\documentclass[aps,pre,reprint]{revtex4-1}

\usepackage{amsmath}
\usepackage{amssymb}

\usepackage{graphicx}

\usepackage{color} 

\date{}

\usepackage{hyperref}

\begin{document}

\title{Functional redundancy in the NF-$\kappa$B signalling pathway}

\author{Micha\l\  W\l odarczyk}
\affiliation{Faculty of Mathematics Informatics and Mechanics, University of Warsaw, ul. Banacha 2
02-097 Warsaw, Poland}

\author{Tomasz\ Lipniacki}
\email[]{tlipnia@ippt.gov.pl}
\affiliation{Institute of Fundamental Technological Research,
Polish Academy of Sciences, ul. Pawi\'nskiego 5B
02-106 Warsaw,
Poland}

\author{Micha\l \  Komorowski}
\email[]{mkomor@ippt.gov.pl}
\affiliation{Institute of Fundamental Technological Research,
Polish Academy of Sciences, ul. Pawi\'nskiego 5B
02-106 Warsaw,
Poland}

\date{\today}
\begin{abstract}
The ability to represent intracellular biochemical  dynamics via deterministic and stochastic modelling is one of the crucial components to move biological sciences in the observe-predict-control-design knowledge ladder. 
Compared to the engineering or physics problems, dynamical models in quantitative biology typically dependent on a relatively large number of parameters. Therefore, the relationship between model parameters and dynamics is often prohibitively difficult to determine. We developed a method to depict the input-output relationship for multi-parametric stochastic and deterministic models via information-theoretic quantification of similarity between model parameters and modules. 
Identification of most information-theoretically orthogonal biological components provided mathematical language to precisely communicate and visualise compensation like phenomena such as biological robustness,  sloppiness and statistical non-identifiability.
A comprehensive analysis of the multi-parameter NF-$\kappa$B signalling pathway demonstrates that the information-theoretic similarity reflects a topological structure of the network. Examination of the currently available experimental data on this system reveals the number of identifiable parameters and suggests informative experimental protocols.\\

{\bf Supplementary Information available at:\\ \indent \url{http://www.ippt.gov.pl/~mkomor/redundancySI.pdf}}\\
\end{abstract}


\maketitle

\noindent  Last decades  accumulated sufficient evidence that a number of biological phenomena, in particular those related to intra-cellular dynamics, noise management, biochemical signalling cannot be understood by intuition alone and require mathematical formalism to explain and summarise  available data. Expectably, mathematical modelling will help in prediction, control and design of biochemical networks.  Therefore adaption of conventional modelling techniques is required to suit  the specificity of these problems. Models of biochemical dynamics are different from classical models of engineering and physics in a number of ways. Primarily they involve substantially larger relative number of parameters compared to available data size.
\noindent This challenge has given rise to a number of approaches aimed at improving our ability to develop, verify and apply multi-parameter mechanistic models of such systems.  We can loosely group these methods into those aimed at  determining model sensitivities to parameter values \cite{Rand2007, brown2003sma, komorowski2011sensitivity}, tools to estimate rate parameters \cite{toni2009approximate,lillacci2010parameter, girolami2011riemann, komorowski2010using,zechner2012moment}, and techniques focused on maximisation of the information content of the experimental data \cite{vanlier2012bayesian,liepe2013maximizing,raue2009structural,kreutz2009systems}.  
The input (parameters) - output (dynamics) dependencies is the main considered object of the above methods. The concept of information, which in the Fisher sense is a sensitivity of an output to parameters, establishes a natural language to communicate a number input-output phenomena. Sensitive parameters exert strong impact on output  and therefore are relatively easy to infer. In consequence, when aiming at parameters estimation, experimental settings,  which render model parameters  sensitive, should be searched.  A number of studies  have reported the intrinsic feature of dynamic multi-parameter models of biochemical dynamics  to be sensitive only to a small number of linear combinations of parameters \cite{brown2003sma,lipniacki2004mmn, Rand2004,Brown2004}. The developed methodology substantial enriched our repertoire of techniques to investigate input-output relationship in multi-parameter models \cite{brown2003sma,gutenkunst2007universally, Rand2007, erguler2011practical, daniels2008sloppiness,hengl2007data,  ludtke2008information, Hengl2007, raue2009structural, komorowski2011sensitivity}. 
\noindent In this paper we built upon these findings to take a comprehensive view at the problem of sensitivities in multi-parameter models. A notion of functional redundancy between individual parameters and their groups (modules) is introduced with Shannon Information being a  measure of its strength. As a result we propose a natural and general mathematical language to precisely communicate and visualise all types of compensation like phenomena i.e. multi-parameter sensitivity, biological robustness,  sloppiness and statistical non-identifiability. It allows for a more insightful interpretation of sensitivity coefficients, detection and elimination of non-identifiable parameters and guided design of  experiments aiming at maximising the number of identifiable parameters. We also find two efficient ways to  evaluate functional redundancy. One is based on the Fisher Information (FI), therefore is local in the parameters space and requires parameter values as input; second is local in the space of experimental results and is based on posterior distribution sampling.  We also integrate the introduced redundancy measure   with a hierarchical clustering algorithm  to informatively represent redundancy structure in form of dendrograms so that functionally related / orthogonal biological components can be easily identified and conclusions about sensitivity, identifiability, robustness can be made.

\noindent The method and its underlying principles are very general and applicable to deterministic and stochastic models. The limiting factor is the computational power. In this paper we focus mainly on the computationally least demanding scenario, deterministic model with available parameter guesses, and report unprecedented insight about parameters redundancies and their consequences for systems biology modelling.  The potential of our framework is demonstrated using an example of the NF-$\kappa$B signalling pathway.  Interestingly, we find that redundant parameters describe modules that are close in the network topology.  We analyse  experimental protocols published in the literature \cite{delhase1999positive, lee2000failure, Hoffmann2002, nelson2004oscillations,werner2005stimulus,lipniacki2007single,werner2008encoding,ashall2009pulsatile,tay2010single}  and find that {\color{black} 26} out of {\color{black} 39} model parameters can be estimated from literature data. In addition, we use our method to propose {\color{black} 10} stimulation experimental protocols, which are expected to provide estimates of {\color{black} 7}  parameters  non-identifiable so far.  In the Supplementary Information (SI) we also calculate redundancies for a JAK-STAT model \cite{vanlier2012bayesian, swameye2003identification}. We consider a stochastic model and a deterministic model with unknown parameter values. 
\section{Functional redundancy  (Methods)}

\begin{figure*}\label{Fig_redundancy}
\begin{center}  \includegraphics[scale=0.72]{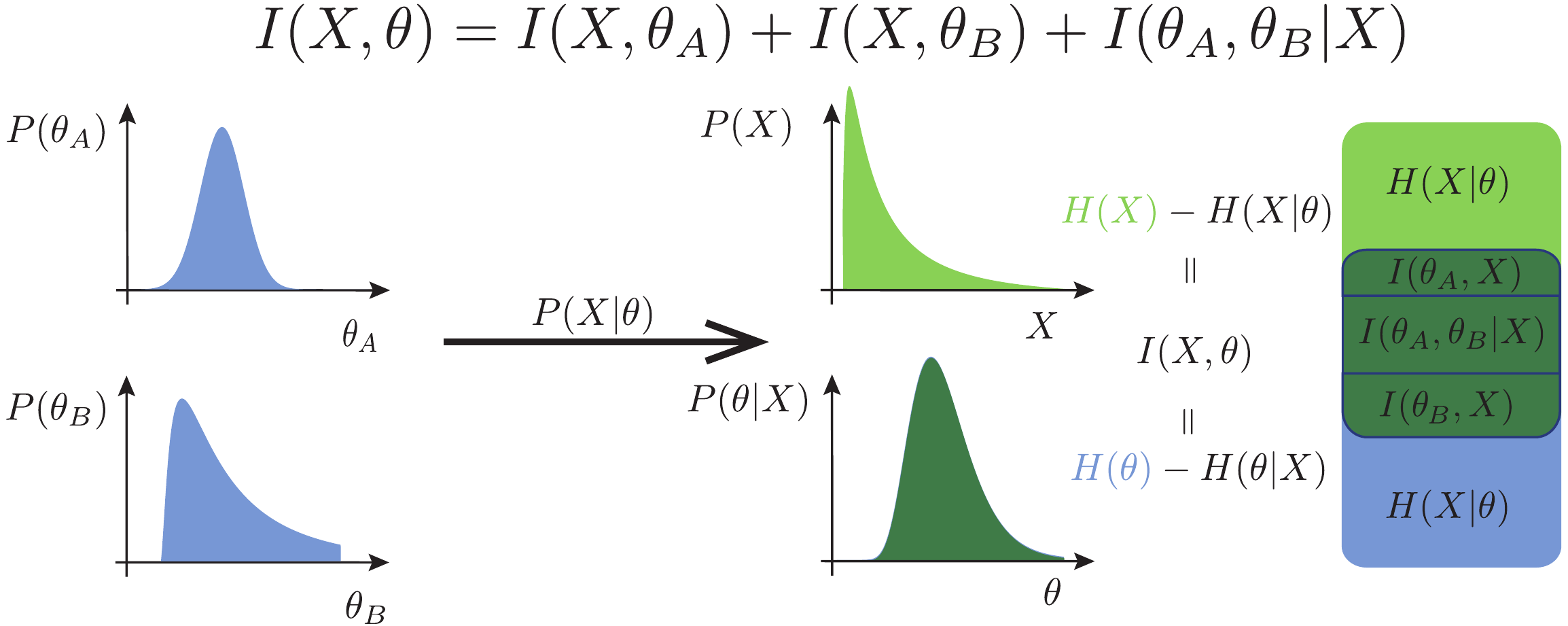}
\end{center}
\caption{
Illustrative description of the functional redundancy. 
The uncertainty about $X$ arises from an uncertainty about parameters $\theta=(\theta_A, \theta_B)$ via model $P(X|\theta)$ and distributions $P(\theta_A)$ and $P(\theta_B)$. The part of the entropy of $X$ that can be explained by $\theta$ is the mutual information $I(X,\theta)$, which arrises from its three constituents: part of the entropy which can be explained solely by $\theta_A$ i.e. $I(\theta_A,X)$; solely by $\theta_B$ i.e. $I(\theta_B,X)$;  and only by concurrent  knowledge of $\theta_A$, $\theta_B$ i.e. $I(\theta_A,\theta_B|X)$. If $\theta_A$ and $\theta_B$ are highly redundant knowing one parameter at a time will not reduce uncertainty about $X$ as it will be reproduced to by the uncertainty about the remaining parameter. Dually, the reduction in uncertainty about $\theta$ resulting from data, $X$, does not directly translates into reduction of individual entropies of $\theta_A$ or $\theta_B$ if these are functionally redundant.
\label{Fig_entropy decomposition} 
}
\end{figure*}
\begin{figure}
\begin{center}  \includegraphics[scale=0.35]{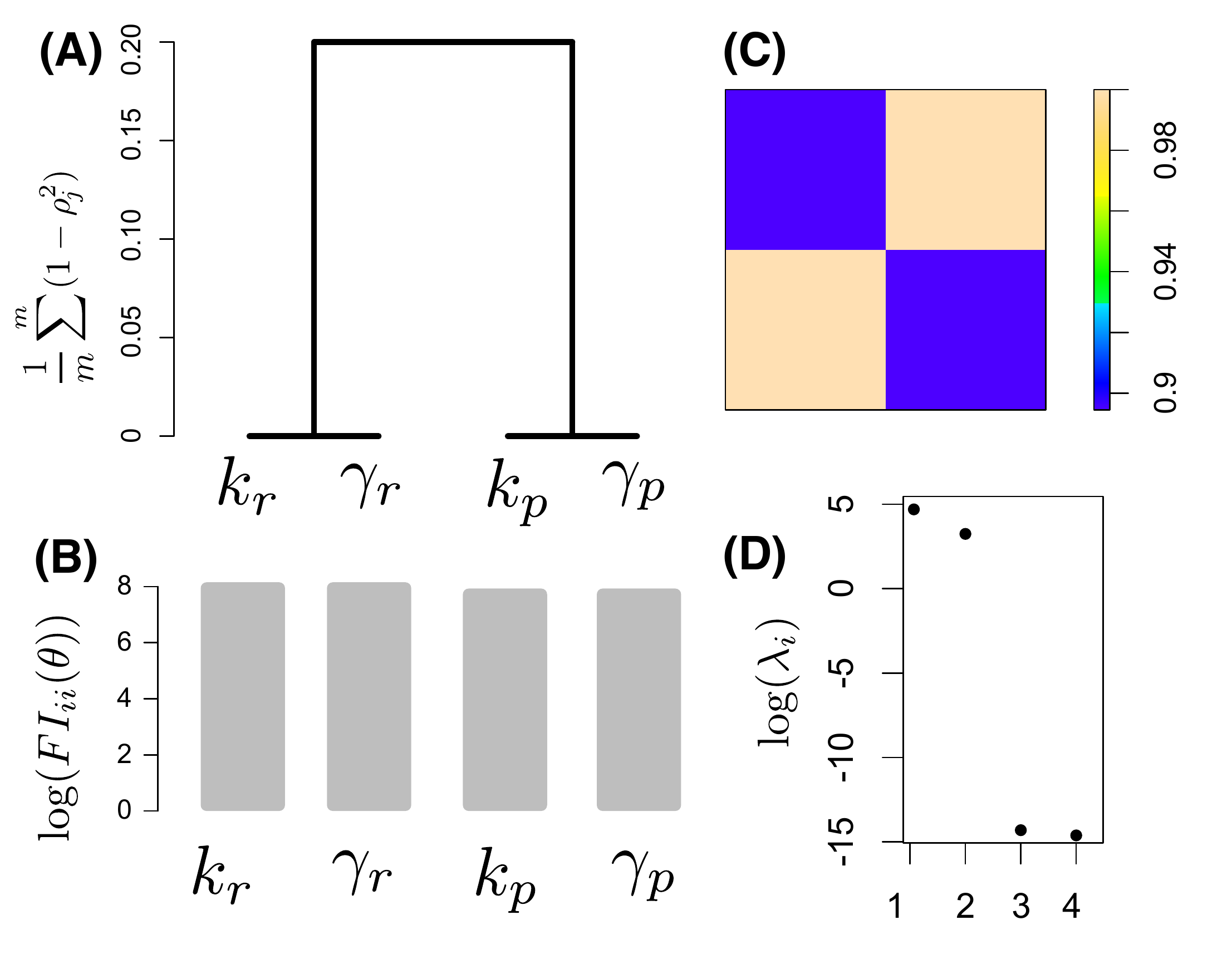}
\end{center}
\caption{Functional redundancy for the simple model of gene expression. {\bf{(A)}} Dendrogram. Linkages are plotted at height    $-\frac{1}{m} \sum_{j=1}^m log( {1-\rho_j^2} )$.  {\bf{(B)}} Sensitivity coefficients: diagonal elements of the FI matrix. {\bf{(C)}} Normalised FI matrix. {\bf{(D)}} Sensitivity spectrum i.e log-eigenvalues of the FI matrix. Parameters used: $k_r= 100$ , $k_p =2$, $\gamma_r=1.2$, $\gamma_p=1$.
\label{Fig_singlegene} }
\end{figure}
\begin{figure}
   \includegraphics[scale=0.39]{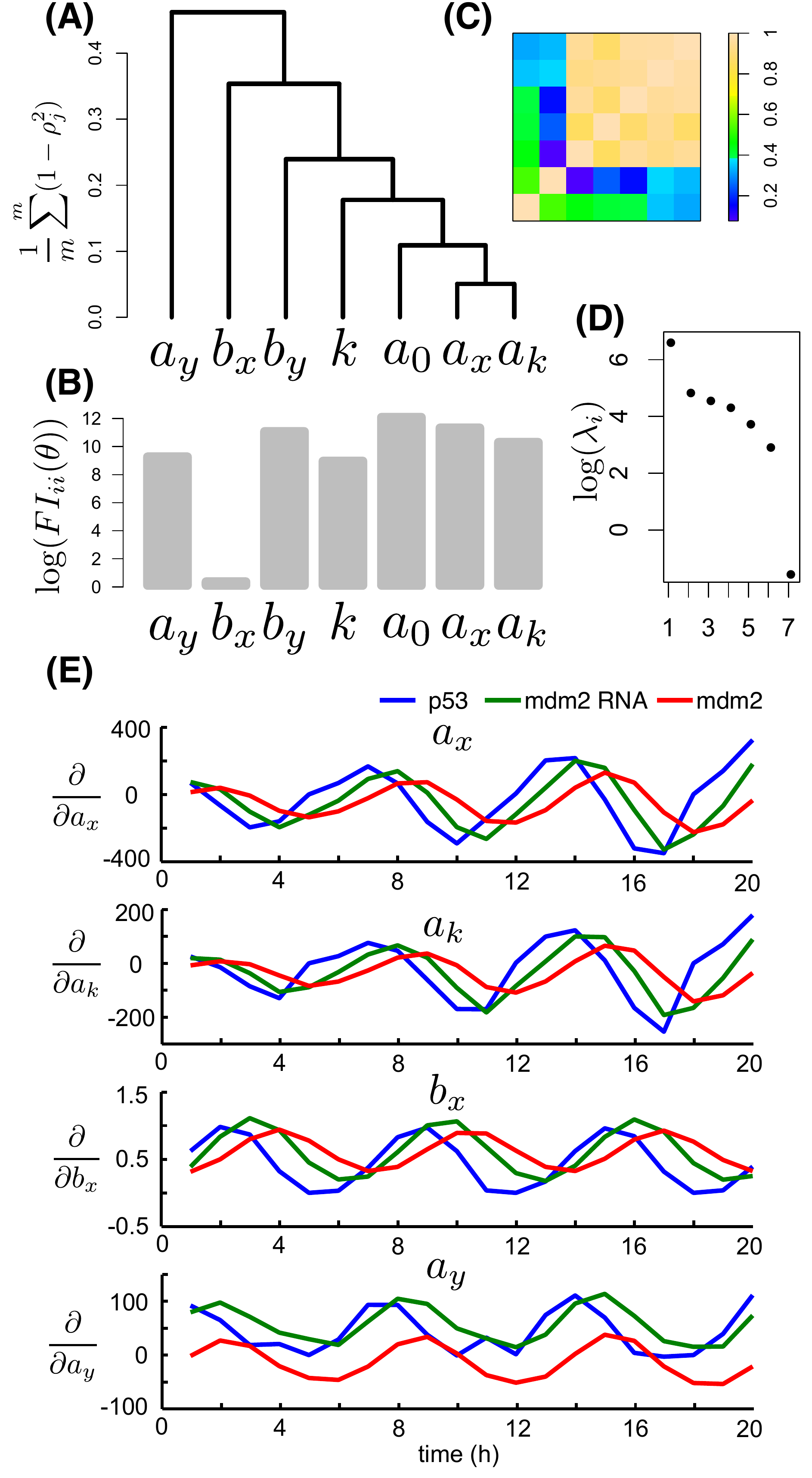} 
\caption{Redundancy analysis of the p53 model. {\bf (A)} Dendrogram. {\bf (B)} Sensitivity coefficients.  {\bf (C)} Normalised FI matrix. {\bf (D)}  Log - eigenvalues of the FI matrix.   {\bf (E)} Derivatives of the model trajectories with respect to selected parameters. Derivatives of $a_0, a_x, a_k$ exhibit almost identical pattern, which is distinct from of this of $a_y$, $b_x$. Parameters used: $\beta_x=90, \alpha_x=.002,\alpha_{k}=1.7,k=0.01,\beta_y=1.1,\alpha_0=0.8, \alpha_y=0.8.$
\label{Fig_p53}  }
\end{figure}

\noindent A typical biochemical kinetics model describes how the abundances of a set of $k$ molecular entities, $y=(y_1,...,y_k)$, change with time, $t$. Most generally the distribution  over times, $(t_1,..., t_n)$, of abundances, $Y=(y(t_1),...,y(t_n))$, can be written as a probability distribution $P(Y|\theta),$ where
$\theta=(\theta_1,...,\theta_l)$ is a vector containing model parameters. Often $y(t)$ is modelled as a solution of ordinary or stochastic differential equation 
\begin{equation}
\frac{dy}{dt}=F(y,\theta)+\xi(t),
\end{equation}
where $\xi(t)$ constitutes a random perturbation in the stochastic setting and  is not included ($\xi(t)=0$) in the deterministic regime.
Practically, only certain components of $y$ can be measured experimentally with certain measurement error $\epsilon$. Therefore we consider variable $X$, which contains selected elements of $Y$ and measurement noise. In the Bayesian setting, the  distribution of data  given parameters, $P(X|\theta)$, together with the prior distribution, $P(\theta)$, define, through averaging over possible parameter values, the distribution of possible measurements, $P(X)$. The uncertainty in  the possible measurements can be formally quantified in terms of the Shannon entropy, $H(X)$. The average reduction in entropy of $X$ gained by knowing $\theta$ is given by the mutual information $I(X,\theta)=H(X)-H(X|\theta)$. The entropy of $X$ therefore can be seen as  resulting from uncertainty in $\theta$ and other sources: $H(X)=I(X,\theta)+H(X|\theta)$.
Assume decomposition of parameters vector $\theta=(\theta_A, \theta_B)$ and set a component-wise independent prior $P(\theta)=P(\theta_A)P(\theta_B)$. The entropy $H(X)$ can now be divided \cite{ludtke2008information} into constituents resulting from 
components of $\theta$
\begin{eqnarray}\label{IXTheta}
H(X) = I(X,\theta_A) +I(X,\theta_B)  &+&  I(\theta_A, \theta_B| X) \\ \nonumber &+& H(X|\{\theta_A,\theta_B \}).
\end{eqnarray}
The reduction of entropy resulting from the sole knowledge of either $\theta_A$ or $\theta_B$ is described by $I(\theta_A)$ and $I(\theta_B)$ respectively.
Mutual information $I(\theta_A,\theta_B|X)$ measures the part of entropy corresponding to the concurrent  knowledge of $\theta_A$ and $\theta_B$. Intuitively, if $\theta_A$ and $\theta_B$ have redundant role knowing one parameter will not reduce uncertainty about $X$ as it will be reproduced by uncertainty in the remaining parameter. Therefore we propose to use  $I(\theta_A,\theta_B|X)$ as a natural measure of redundancy between model parameters. The redundancy has also a dual interpretation. The reduction in uncertainty of $\theta$ gained by observing data $X$ decreases the entropies of  $\theta_A$ and $\theta_B$ to a degree dependent on their redundancy level measured by $I(\theta_A,\theta_B|X)$. See Fig. \ref{Fig_redundancy} for illustration and SI for technical details.\\
Evaluation of $I(\theta_A, \theta_B| X)$ in a general setting it is computationally prohibitively expensive for most models of realistic size (see SI for details).  We show, however, how to evaluate $I(\theta_A, \theta_B| X)$ efficiently in two relevant scenarios. In the first one,  we assume that guesses about true parameter values are available, in the second that experimental data can be used to generate posterior distribution. Denote available parameter guesses as $\theta^{*}$. In this case the posterior distribution can be approximated using the bayesian asymptotic theory 
\begin{equation}
P(\theta|\theta^{*})\propto \exp( -\frac{1}{2N}(\theta-\theta^{*} )FI(\theta^{*})(\theta-\theta^{*} )^T ),
\end{equation}
where $FI$ is the Fisher Information matrix  of the model $P(X|\theta^{*})$ and $N$ is the number of replicates of $X$ considered in the above posterior. If $FI$ has the full rank then the above density describes the multivariate normal distribution (MVN). In this case we found a straightforward and numerically stable method  to evaluation mutual information
\begin{equation}\label{AMI_CC}
I(\theta_A, \theta_B|\theta^{*})=-\frac{1}{2}\sum_{j=1}^m \log( {1-\rho_j^2} ),
\end{equation}
where $\rho_j$ are canonical correlations extracted from the $FI(\theta^{*} )$ and $m$ is a minimum of lengths of $\theta_A$ and $\theta_B$ (see SI for derivation).  Canonical correlations are defined as the maximal correlation between mutually orthogonal linear combinations of parameters of $\theta_A$ and $\theta_B$ (see SI).
Therefore we can easily quantify functional redundancy if FI can be computed. This can be efficiently done, virtually without any computational limitations, for deterministic models and for models of moderate size in the stochastic setting \cite{komorowski2011sensitivity}. 
The second computationally tractable case is  the one with available experimental data, $X^{*}$. Here mutual information $I(\theta_A, \theta_B| X=X^{*})$ can be estimated based on a sample from a posterior distribution $P(\theta|X^{*})$. Sampling can be performed using one of the available Monte Carlo approaches (see SI for details). We show that despite limitations to evaluate computational redundancy in a general setting an unprecedented insight can be gained by computing redundancy in the two above scenarios. For simplicity, from now on we write $I(\theta_A, \theta_B)$  instead of $I(\theta_A, \theta_B|\theta^*)$ and $I(\theta_A, \theta_B|X^*)$.\\
{\it Clustering to reveal functional parameters redundancy.}
Being able to evaluate $I(\theta_A, \theta_B)$ we can represent redundancy structure for a given model using  statistical clustering tools. Among a number of clustering approaches \cite{slonim2005information, friedman2001elements} we chose the hierarchical clustering, which is more appropriate to represent correlation structure with assignment to clusters having a secondary importance (see SI for details).  In order to built an intuition how a model can be analysed using this approach we start with the toy model of gene expression. \\
{\it Gene expression.} Assume that the gene expression process begins with the production of mRNA molecules (r) at rate $k_r$. Each mRNA molecule may be independently translated into protein molecules (p) at rate $k_p$. Both mRNA and protein molecules are degraded at rates $\gamma_r$ and $\gamma_p$, respectively. We consider steady state behaviour of deterministic model.  Therefore, we have the state vector $(r,p)=\left(\frac{k_r}{\gamma_r},\frac{k_r k_p}{ \gamma_r \gamma_p} \right)$. This formula suggests that parameters  $k_r$ and $\gamma_r$ are entirely redundant as is the pair $k_p$ and $\gamma_p$.  Parameters pairs $(k_r, \gamma_r)$ and $(k_p, \gamma_p)$ are only partially redundant as the ratio $k_r / \gamma_r$ has identical impact on protein level as the ratio $k_p / \gamma_p$ and the latter does not impact mRNA level. This redundancy, reproduced by clustering algorithm, is visualised by the dendrogram  in Figure \ref{Fig_singlegene}. Parameters $k_r, \gamma_r$ and $k_p, \gamma_p$ are first linked reciprocally at 0 hight and both pairs are linked together at non-zero high.  We plot the linkages at height  $-\frac{1}{m} \sum_{j=1}^m \log( {1-\rho_j^2} )$, (where $m$ is a size of a smaller cluster) compared to the maximum of sons height. Canonical correlation gives a clear interpretations to linkages heights, which is between 0 and 1.  The heatmap (top left corner) is the normalised FI matrix, $|\frac{FI_{ij}(\theta^{*})}{\sqrt{FI_{ii}(\theta^{*}) FI_{jj}(\theta^{*})}}|$. Sensitivity coefficients, $FI_{ii}(\theta)$, are plotted in the bottom left corner.  The steady state behaviour of this simple system is robust to  mutually compensating perturbations of production and degradation rates. The compensation renders two parameters identifiable, one of each of the pairs $(k_r, \gamma_r)$ and $(k_p, \gamma_p)$. This redundancies can be removed by manipulating the initial condition and observing system temporal dynamics. \\
{\noindent\it The  p53 system.} 
In the above simple linear model relationship between behaviour and parameters can be understood well without our information theoretic analysis.  In a more complex, though still simple example, model of the p53 signalling system,  a feedback loop between the tumor suppressor p53 ($x_p$), the oncogene Mdm2  mRNA ($x_0$) and Mdm2 ($x_1$), generates oscillation in response to DNA damage. Hence, the relations between parameters is rather obscure. The deterministic version of the model is formulated as follows  \cite{geva2006oscillations}
\begin{eqnarray}
\dot{x}_p&=& \beta_x-\alpha_x x_p -\alpha_{k}x_y\frac{x_p}{x_p+k}\\\nonumber
\dot{x}_{0}&=&\beta_y x_p- \alpha_0 x_{0} \\\nonumber
\dot{x}_1&=& \alpha_0 x_{0} - \alpha_y x_{1}.
\end{eqnarray}
Our approach can  provide a comprehensive qualitative summary of parameters redundancy given that the parameter values  are provided. We use values published in \cite{geva2006oscillations}.  The dendrogram in Fig. \ref{Fig_p53}A predicts that of all parameters 
$\theta=( \beta_x, \alpha_x,\alpha_{k},k,\beta_y,\alpha_0, \alpha_y )$ the  $a_0, a_x$ exert most similar impact on model dynamics and  $b_x,a_y$ are 'least similar' ones. This is confirmed by plotting derivatives of model trajectories (Fig. \ref{Fig_p53}{B}) with respect to these parameters. Conclusions regarding model robustness and identifiability can be easily analogised  to the previous example. 
\section{Results}
\begin{figure*}
 \includegraphics[scale=0.25]{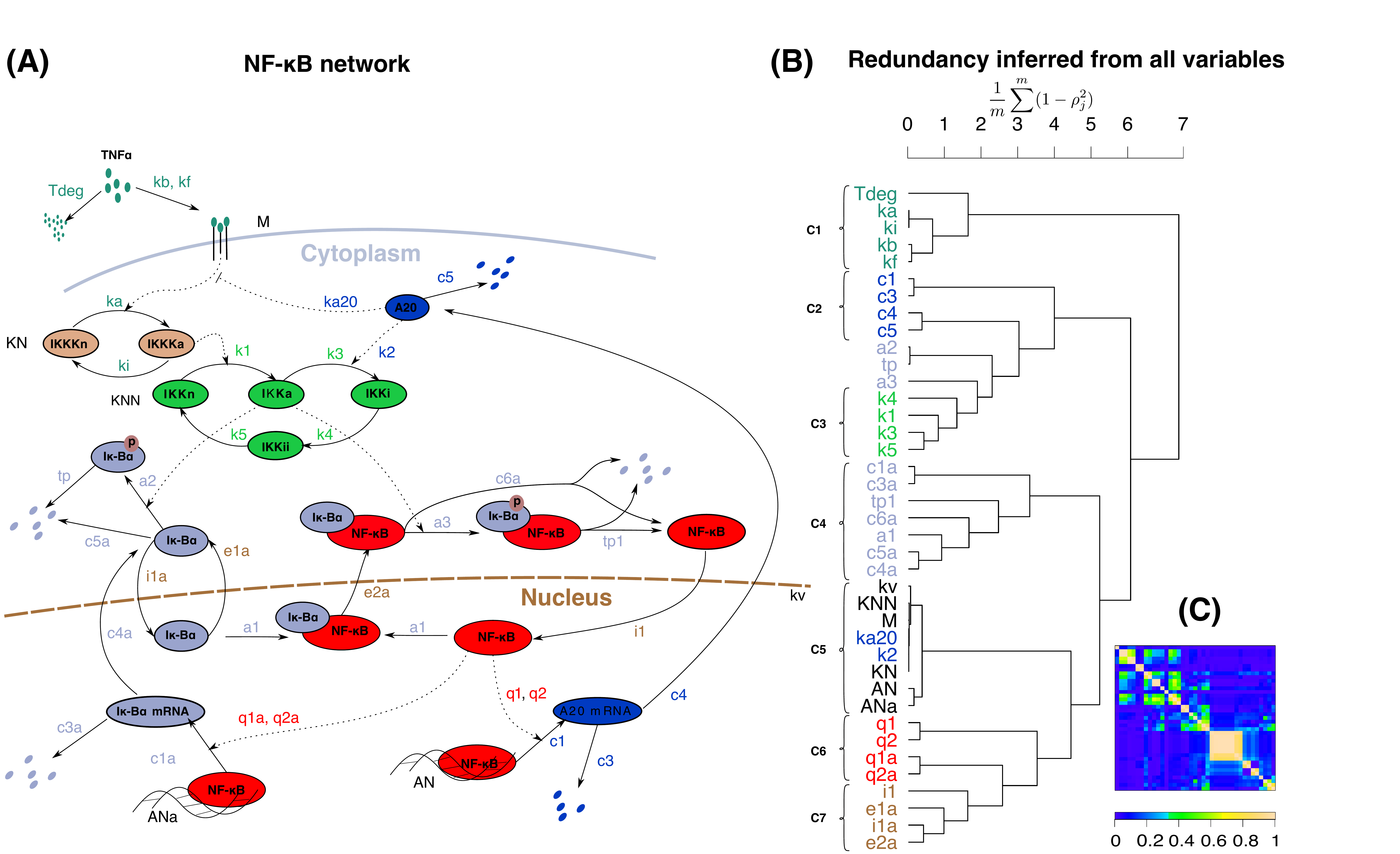}
\caption{ Functional redundancy in the 
NF$\kappa$B signalling pathway. {\bf (A)} Structure of the network.    {\bf (B)} Dendrogram of functional redundancy represented as dendrogram. Structure was computed assuming all model variables normalised by their maxima are observed with normal zero mean, unit standard deviation measurement error. Parameters were colour coded to represent biochemical elements of the network. Specific TNF-$\alpha$ stimulation to reflect physiological conditions (increase, plateau, decrease) was applied (see Fig. 1B in SI). Clusters $C1-C8$ exhibit good correspondence with functional components of the network. A corresponding analysis under the assumption that the nuclear NF-$\kappa$B trajectory is the only observable is presented in the Figure 1A in SI.  {\bf (C)} Normalised Fisher Information Matrix used to construct dendrogram. 
\label{Fig_NFkBtop} }
\end{figure*}

\begin{figure*}
 \includegraphics[scale=0.35]{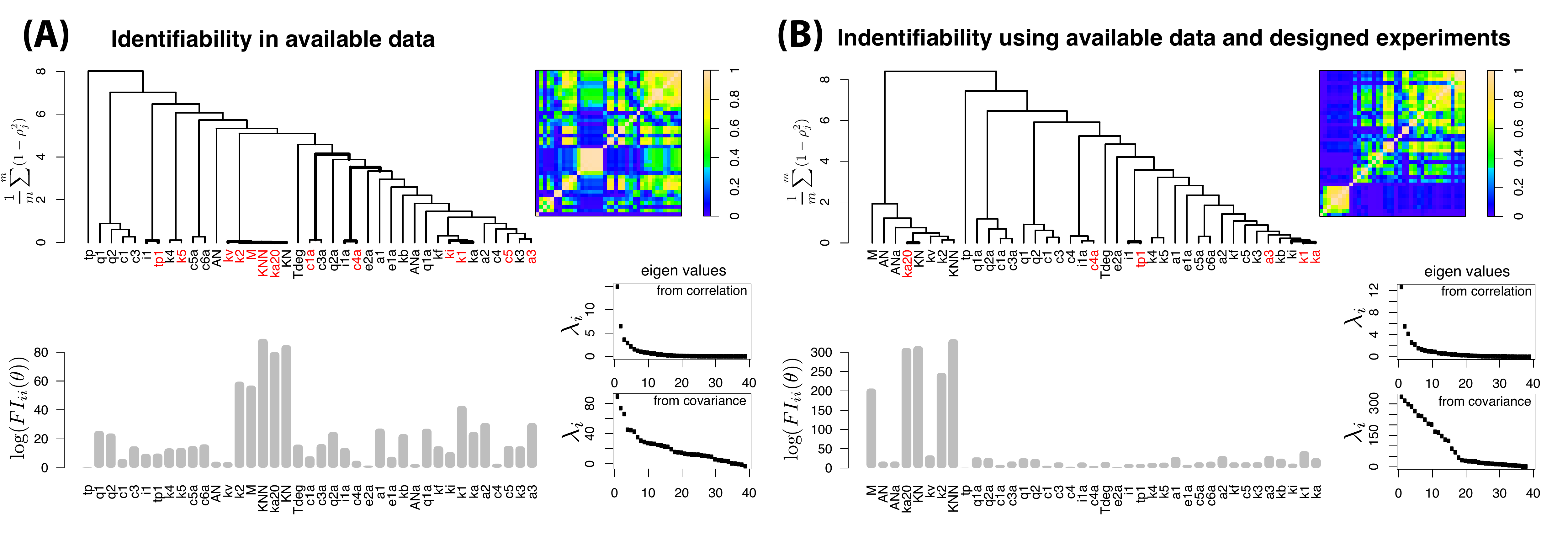}
\caption{{\bf (A)} Redundancy structure indicates identifiable parameters given experimental data published in  \cite{delhase1999positive, lee2000failure, Hoffmann2002, nelson2004oscillations,werner2005stimulus,lipniacki2007single,werner2008encoding,ashall2009pulsatile,tay2010single}. Setting  $\delta=0.05$, $\zeta=1$ we found that 26 parameters can be identified (plotted in black, the non-identifiable parameters in red.  Sensitivity coefficients i.e diagonal elements of FI matrix are shown below the dendrogram. The heatmap plot presents the normalised FI matrix, its eigen spectrum is shown below. Linkages plotted in bold violate the condition $\rho(\theta_i, \theta_{-i})<1-\delta$.
 {\bf (B)} Same as in (A) but for the experiments published  in \cite{delhase1999positive, lee2000failure, Hoffmann2002, nelson2004oscillations,werner2005stimulus,lipniacki2007single,werner2008encoding,ashall2009pulsatile,tay2010single} together with 10 best experimental protocols (Fig.  \ref{Fig_protocols}) found in a random search.
\label{Fig_NFkBexpdata} }
\end{figure*}

\begin{figure}[h]
\begin{center}  \includegraphics[scale=0.35]{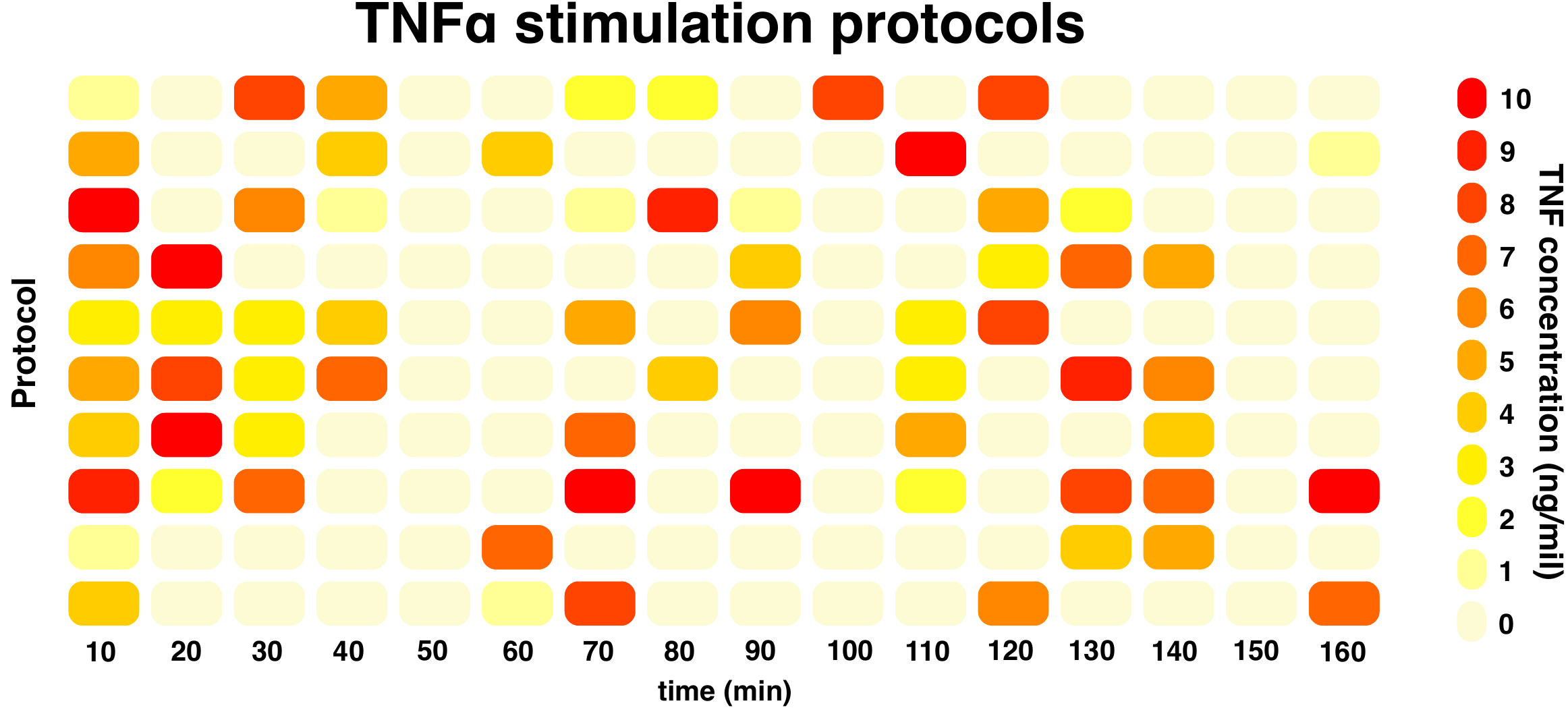}
\end{center}
\caption{TNF-$\alpha$ stimulation protocols designed to increase the number of identifiable parameters.  Each row describes one of the 10 found temporal profiles
The protocols together with data of  \cite{delhase1999positive, lee2000failure, Hoffmann2002, nelson2004oscillations,werner2005stimulus,lipniacki2007single,werner2008encoding,ashall2009pulsatile,tay2010single}
allow to estimate 7 new parameters compared to using data of \cite{delhase1999positive, lee2000failure, Hoffmann2002, nelson2004oscillations,werner2005stimulus,lipniacki2007single,werner2008encoding,ashall2009pulsatile,tay2010single} only. Details of the protocols are presented in  the {\color{black}  Table 4 in the SI} .\label{Fig_protocols} }
\end{figure}

{\noindent \it  The NF-$\kappa$B system. } 
Here we demonstrate the potential of our method using a deterministic model of the NF-$\kappa$B signalling pathway, one of the key components  controlling innate immune response. In particular we address four important problems: $1)$ identify most information-theoretically orthogonal components of the network that control its behaviour;  $2)$ characterise robustness properties of the pathway;
$3)$ analyse all published experimental protocols to asses which parameters can be estimated from the available data; $4)$  propose stimulation protocols, which can increase the number of identifiable kinetic constants. The model considered here, proposed in \cite{lipniacki2004mmn}  and further developed in \cite{tay2010single} represents activation of NF-$\kappa$B induced genes in response to  stimulation by TNF-$\alpha$ (the pro-inflammatory cytokine). It involves 39 parameters and 19 variables and encapsulates typical features of systems biology models that challenge current modelling techniques.\\
{\it  Redundant control of the response to TNF-$\alpha$ stimulation.}
Visualisation of the redundancy structure enables identification of network components that contain most redundant parameters and those  most orthogonal that control network dynamics.  We first analysed scenario where behaviour of the system, $X$, is defined by trajectories of all model variables normalised by  their maxima measured with normal measurement error  $\epsilon \sim MVN(0,I)$ ($X_i=Y_i/max(Y_i)+\epsilon$). We considered reminiscence of physiological TNF-$\alpha$ stimulation profile and  assumed TNF-$\alpha$ concentration to increase and drop after an intermediate plateau (see Figure {\color{black} 1B of SI}).   The constructed redundancy dendrogram (Fig. \ref{Fig_NFkBtop}) interestingly indicates that redundant parameters are grouped into clusters, denoted by $C1$-$C7$, that to a large extent correspond to their topological localisation. Cluster $C1$ contains parameters describing receptor activation and signalling; $C2$: A20 synthesis and degradation; $C3$: IKK kinase post-translational modifications, I$\kappa$B$\alpha$ synthesis, degradation, phosphorylation, and interaction with IKKK and NF-$\kappa$B; $C4$: I$\kappa$B$\alpha$ synthesis, degradation and interaction with NF-$\kappa$B; $C5$: molecule numbers and cell characterisation (except $ka20$ and $k2$); $C6$: NF-$\kappa$B - DNA interactions;  $C7$: nuclear shuttling. We can also consider  two bigger clusters, one composed of elements of $C1$ and $C2$, and the second of ($C3$-$C8$), 
which correspond to the external feedback loop controlled by A20, 
and internal feedback loop controlled by primary inhibitor IkBa, respectively. Parameters within the modules have a similar impact on model dynamics and their perturbations can be compensated by changes of other parameters within the module. Module by module compensation is less efficient i.e. correlation (mutual information) between clusters is smaller than between parameters within the clusters.\\
\noindent  As a second scenario we consider the behaviour of the system, $X$, to be defined entirely by the normalised nuclear concentration of the NF-$\kappa$B with measurement error. The corresponding dendrogram is presented in Figure {\color{black} 1A of SI}.   The redundancy between parameters is much stronger in this case (many low linkages) and the structure appears to be more random what reflects much lower information content of  measurements.\\
{\it  Robustness properties of the pathway.} The dendrogram in Figure \ref{Fig_NFkBtop} gives a good characterisation of robustness of the response to TNF-$\alpha$ stimulation. In the first case (all variables define model behaviour)  robustness is a topologically-local property i.e. system is robust to perturbations that compensate each other locally. In the second scenario (nuclear NF-$\kappa$B defines model behaviour), the system is robust to a much wider class of perturbations as most of the parameter pairs are mutually compensative. Below we also show that redundancy structure is stimulus/experiment specific.  Stimuli that break functional redundancy are desired to infer model parameters from experimental data and we show that it is possible to design stimuli which reduce the redundancy. \\
\noindent {\it  Redundancy and identifiability in available experimental data.}
Highly redundant parameters have almost identical impact on observed experimental data and therefore render  non-identifiability. There is a number of formal definitions of identifiability \cite{raue2009structural,rothenberg1971identification, audoly2001global}. Parameters are structurally non-identifiable if the system exhibits  identical behaviour for two different values of these parameters. Locally,  this can be detected solely based on the FI. Structurally identifiable parameters can be practically non-identifiable when for a given experimental data set likelihood is almost flat with respect to these parameters \cite{raue2009structural}. In our setting this corresponds to the case where  mutual information is high (alternatively $\rho$ is close to 1) or the corresponding sensitivity coefficient is small. The existing definitions together with our  approach motivate to define identifiability in terms of redundancy. We introduce definition of ($\delta$, $\zeta$)-identifiably. We say that an element  $\theta_i$ of the vector $\theta=(\theta_1,...,\theta_k)$ is ($\delta$,$\zeta$)-identifiable if  $\rho(\theta_i, \theta_{-i})<1-\delta$ (equivalently $I(\theta_i, \theta_{-i})<\log(1/\sqrt{\delta(2-\delta)})$)  and $FI_{ii}(\theta)>\zeta$ (see SI), where $\theta_{-i}$ denotes all elements of the vector $\theta$ except $\theta_i$. This definition is rooted in the conventional statistics. If all other parameters were known, the standard deviation of  a most efficient estimator of $\theta_i$, $sd(\theta_i)$, is given by the asymptotic formula $sd(\theta_i)=1/{\sqrt{FI_{ii}(\theta)}}$. The $\zeta$ requirement  demands the individual standard  deviations to be smaller then $1/{\sqrt{\zeta}}$.  The $\delta$ condition requires the asymptotic standard deviation not to increase more than 
$(1/{\sqrt{\delta(2-\delta)}})$-fold when all elements of $\theta$ are estimated at once compared to the previous case (see SI for detailed explanation). Using the constructed criteria we asked how many parameters of the model \cite{lipniacki2004mmn, tay2010single} can be estimated from data available in the literature.   We select $9$ papers that contain rich data sets on the dynamics of the NF-$\kappa$B system \cite{delhase1999positive, lee2000failure, Hoffmann2002, nelson2004oscillations,werner2005stimulus,lipniacki2007single,werner2008encoding,ashall2009pulsatile,tay2010single}, which could be used for parameter inference. All  experimental measurements are summarised in Table 1 in SI.  We arbitrarily set $\delta=0.05$, which corresponds to the increase of the asymptotic variance less than 10 times;  and $\zeta=1$. As we use logarithms parametrisation, i.e. $\log(\theta_i)$ instead of $\theta_i$,  setting $\zeta=1$  corresponds to learning a parameter with an order of magnitude error, if all other parameters were known.   Under this assumptions we found that {26} parameters can be estimated. The identifiable parameters are plotted black in the Figure \ref{Fig_NFkBexpdata}.
Among non-identifiable parameters we found  $M, KN, KNN$ describing levels of receptors, IKKK kinase and IKK kinase respectively and parameters $k2$ and $ka20$ describing  signalling mediated by receptors, IKKK and IKK.\\
{\it  Informative future experiments.} The redundancy structure for all experiments  of \cite{delhase1999positive, lee2000failure, Hoffmann2002, nelson2004oscillations,werner2005stimulus,lipniacki2007single,werner2008encoding,ashall2009pulsatile,tay2010single}  presented in Fig. \ref{Fig_NFkBexpdata} indicates that certain parameters cannot be inferred as a result of their redundancy with other parameters of the model. Non-identifiabilty does not result from parameters being insensitive individually.  To find out if and to what extent the redundancy can be eliminated by designed stimuli we randomly searched a space of potential new stimuli (see SI) and selected 10 experiments that collectively give the highest number of identifiable parameters.  To ensure practical value of our guidance we assumed that only the following entities were measured: $I\kappa$B$\alpha$ protein (blotting), $I\kappa$B$\alpha$ mRNA,   nuclear NF-$\kappa$B (fluorescence microscopy), and activity of IKK. The measurements were assumed to be taken every 5 minutes for 160 minutes with normally distributed error with standard  deviation equal to the squared root of  possible maximum of the measurement.
{ We found that by caring out 10 experimental stimulations with TNF-$\alpha$ presented in Fig. \ref{Fig_protocols} and Table {\color{black} 4 in the SI} we can identify values of {\color{black}7} new parameters. New experiments involve pulses with frequency ranging between $1/4 - 1/2$ of the NF-$\kappa$B oscillation frequency which is approximately 100 min. Such protocols have not been performed so far.}

\section{Discussion}
Tools to understand the relationship between parameters and model dynamics seem to be of high relevance to utilise the potential of mathematical modelling within bimolecular sciences. The complexity of dynamic quantitative biology models makes their manipulation a subject of time consuming and laborious investigations. It is so because components and parameters are not independent (orthogonal) but exert their impact jointly through the network of interactions.  A number of studies reported on the analysis of multi-parameter models.
A notion of  the sloppy model was introduced in \cite{brown2003sma, Brown2004,gutenkunst2007universally}  to describe the model property of being sensitive only to a small number of linear combinations of parameters. New technique for sensitivity analysis   that takes into account varying contributions of parameters into sensitivity spectrum was proposed in \cite{Rand2007}. Information theory was applied in \cite{ludtke2008information} to  reveal higher order  interactions between model parameters.  Novel methods  proposed in  \cite{Hengl2007, raue2009structural} can identify linear and non-linear relations between parameters and detect  non-identifiability in experimental data. 
\noindent A conjectured solution to the problem of over-parametrized models is to find and manipulate orthogonal model components. This however creates abstract objects that may not have a relevant biological interpretation.  Our method enables  visualisation  of similarities (redundancies) to find most redundant and most orthogonal biological components. It constitutes a unique mathematical framework to describe  a variety of compensation like phenomena. Generality of the method enables its applicability to stochastic and deterministic models (see SI). A most detailed insight is provided if model parameters are known. Nevertheless, information about redundancies can also be inferred directly from experimental data (see SI). The accuracy of inferred similarities depends on quality of available data. \\
\noindent In the paper we focused on the computationally least demanding scenario of deterministic models with available parameter estimates. We demonstrated the potential of our method by addressing relevant questions pertaining to the dynamics of the NF-$\kappa$B system. We shown that functionally related parameters are topologically co-localised. Modules formed by redundant parameters impact the dynamics of the system in a more independent manner. This has consequences for model robustness and parameter identifiability, particularly implies that it is most difficult to infer parameters that are close in the networks topology. We also examined the literature available experimental protocols to show how redundancy disables inference of parameters of the NF-$\kappa$B dynamical model and suggested TNF-$\alpha$ stimulation protocols to break the redundancy and  infer more parameters. In the SI we shown how  the method can be applied to analyse stochastic models of moderate size.  Utilisation of the method for scenarios where only vague priors on parameter values are available requires further development of computational techniques. \\
\noindent The introduced concept of redundancy shows how components of biological systems interrelate exerting the joint impact on observed biochemical dynamics. Therefore it has a tangible potential to overcome some of the difficulties resulting from the complexity of models in quantitative biology.
\begin{acknowledgments}
MW and MK were supported by the Foundation for Polish Science under
the program Homing Plus HOMING 2011-3/4. TL acknowledges support from Foundation for Polish Science under
the program TEAM 2009-3/6. MK is also EMBO Installation Grantee.  We thank Marek Kocha\'nczyk for his valuable comments on this work. 
\end{acknowledgments}\ \\

\end{document}